\documentclass[preprint,authoryear,12pt]{elsarticle}

\usepackage{graphicx}
\usepackage{bm}
\usepackage{amsmath}
\usepackage{natbib}
\usepackage{graphicx}
\usepackage{comment}
\usepackage[colorlinks=true]{hyperref}   

\usepackage{leponcin}
\usepackage{ads}

\usepackage{todonotes}

\newcommand*{\norm}[1]{\mathopen\| #1 \mathclose\|}

\def\norm#1{\mathopen\| #1 \mathclose\|}

\usepackage{epsfig}

\usepackage{amssymb}

\journal{Advances in Space Research}

\def\bx{{\bm x}}

\def\bN{{\bm N}}
\def\be{\begin{equation}}
\def\ee{\end{equation}}
\def\bea{\begin{eqnarray}}
\def\eea{\end{eqnarray}}
\def\lb{\label}
\def\nn{\nonumber}

\begin{document}

\begin{frontmatter}



\title{Impact analysis of the transponder time delay on radio-tracking observables}


\author{Stefano Bertone [1]}\ead{stefano.bertone@aiub.unibe.ch}
\author{Christophe Le Poncin-Lafitte [2]}
\author{Pascal Rosenblatt [3]}
\author{Val\'ery Lainey [4]}
\author{Jean-Charles Marty [5]}
\author{Marie-Christine Angonin [2]}

\address{[1] Astronomical Institute, University of Bern, Switzerland}
\address{[2] SYRTE, Observatoire de Paris, PSL Research University, CNRS, Sorbonne Universit\'es, UPMC Univ. Paris 06, LNE, 61 Avenue de l'Observatoire, 75014 Paris, France}
\address{[3] Royal Observatory of Belgium, Brussels, Belgium}
\address{[4] IMCCE, Observatoire de Paris, CNRS/UMR 8028, 61 Av. de l'Observatoire, F75014 Paris, France}
\address{[5] CNES/GRGS, OMP 14 avenue \'Edouard Belin 31400 Toulouse,
France}


\begin{abstract}

Accurate tracking of probes is one of the key points of space exploration. Range and Doppler techniques are the most commonly used. In this paper we analyze the impact of the transponder delay, $i.e.$ the processing time between reception and re-emission of a two-way tracking link at the satellite, on tracking observables and on spacecraft orbits. We show that this term, only partially accounted for in the standard formulation of computed space observables, can actually be relevant for future missions with high nominal tracking accuracies or for the re-processing of old missions. We present several applications of our formulation to Earth flybys, the NASA GRAIL and the ESA BepiColombo missions.

\end{abstract}


\end{frontmatter}

\parindent=0.5 cm


\section{Introduction}

Accurate tracking of probes is one of the key points of space exploration. Several radio tracking strategies are possible to determine the trajectory of interplanetary spacecraft, but Doppler and Range techniques are the most commonly used. Precise orbits are then the basis of many scientific applications, from geodesy and geophysics to the study of planetary atmospheres, the correct interpretation of instrument data up to fundamental physics experiments.

Range accuracy improved by one order of magnitude during the last 10 years (from $1$ meter for the NASA Cassini probe - \cite{2014PhRvD..89j2002H} - to $10$~cm for the ESA BepiColombo mission - \cite{2001P&SS...49.1579M,Genova_MemSAI}) while, thanks to the development of X- and Ka-band transponders, Doppler accuracy increased drastically from $\approx 10$~mHz for Pioneer Venus Orbiter~\citep[$\approx 1.5$ mm/s @ $2.2$ GHz, ][]{1993GeoRL..20.2403K} 
to the $\mu$Hz level for BepiColombo and for Juno~\citep[$\approx 35$ nm/s @ $8.4$ GHz, ][]{2017AJ....154....2G}.
Improvements in the technical accuracy of these observables result in better constraints on their scientific interpretation and have consequences in several domains. For this reason, a continuous effort is necessary to keep up the modeling with the increasing accuracy of instruments and mission goals. 


In both Range and Doppler techniques, tracking signals are exchanged between an antenna on Earth and the probe.
The standard light-time formulation by~\citet{Moyer:2000} accurately describes how to model this exchange and the resulting observables. 
However, the small time delay between the reception of the tracking signal on the probe and its re-emission back to Earth is currently neglected in the two-way Doppler formulation, while it is introduced as a simple calibration in the two-way range~\citep{MontenbruckGill.book,Moyer:2000}. This time delay, which we call "transponder delay", represents the response time of the transponder electronics and it is around several $\mu$s for modern transponders~\citep{BussoPriv}.

In this paper, we analyze the impact of including this term in the mathematical formulation of computed light-time and deep space Range and Doppler observables with the goal of improving the agreement between computed and observed quantities in the orbit determination process.
Section~\ref{sec:standard_dsn} briefly summarizes the standard modeling as given by~\citet{Moyer:2000}. Then, in Section~\ref{sec:improvedmodel} we describe the introduction of the transponder delay in light-time modeling and in Section~\ref{sec:observables} its impact on the range and Doppler observables. Section~\ref{sec:applications} provides some examples of the impact of the additional terms in several configurations such as an Earth swing-by, NASA GRAIL~\citep{2013Sci...339..668Z} and ESA BepiColombo missions. Finally, in Section~\ref{sec:conclusions} we summarize our final remarks.

\section{Standard modeling of light time for radioscience observables}\label{sec:standard_dsn}


\par The standard approach for space navigation is presented in \citet{Moyer:2000}. It provides the formulation for observed and computed values of deep space navigation data. Only a cursory description is provided as required by the scope of this paper.

Orbit Data File (ODF) tracking data consist of time series of observed Range or Doppler counts. Both these observables can be computed as functions of the time of flight of the signal between the observing station and the probe, provided auxiliary information, $e.g.$, Doppler emitted frequency and count times, are available in the ODF. 
\citet{Moyer:2000} designates the transmission time from Earth of the up-leg link as $t_1$, the epoch of reception and immediate re-transmission as $t_2$  and finally the reception time of the down-leg link on ground as $t_3$. At each of these epochs the Solar System
Barycentric (SSB) position vectors of the up-link station $\bm x_1$, spacecraft $\bm x_2$ and down-link station $\bm x_3$ must be calculated.

All coordinates presented in this paper, unless differently stated, are defined in the Barycentric Reference System~\citep[BCRS, ][]{2010ITN....36....1P}, while all epochs are consistently given in the Barycentric Dynamical Time~\citep[TDB, ][]{2010ITN....36....1P}.
Moreover, we shall neglect all transformations between coordinate and proper times, since the relative modification would be at most $10^{- 8}$, which is fully negligible.



Range observables are related to the distance between observer and receiver, while Doppler observables (in the Moyer's sense of instantaneous Doppler shifts averaged over a time interval $T_c$) provide a constraint on their relative radial velocity.
These so called {\it computed observables} are then used in the orbit recovery process by means of a least square fit to the tracking observations .

\subsection{Round-trip light time} \lb{sec: standardLT}


The key point for the {\it computed observables} is to properly describe the round trip time of flight $\rho$ of the light signal. The standard formulation by~\cite{Moyer:2000} gives
\begin{subequations} \label{eq:moyerLTit}
	\begin{eqnarray} \label{eq:moyerLT}  
		\rho & = & \frac{R_{12}}{c} + \frac{R_{23}}{c} + RLT_{12} + RLT_{23} + \delta \rho 
	\end{eqnarray}
where 
	\begin{eqnarray}
		\frac{R_{12}}{c} &=& t_2(ET) - t_1(ET) = \frac{\norm{\bx_2 - \bx_1}}{c} \; , \\
	    \frac{R_{23}}{c} &=& t_3(ET) - t_2(ET) = \frac{\norm{\bx_3 - \bx_2}}{c} \; ,
	\end{eqnarray}
\end{subequations}
and $RLT_{ij}$ is the Shapiro delay~\citep{1964PhRvL..13..789S} on the up-leg and down-leg light time solutions. 
Moreover, we noted $\delta\rho$ the additional delays ($e.g.$, atmospheric and instrumental delays at ground stations) and ET the ephemeris time. 
Since modern ephemeris also define the TDB consistently with planetary ephe\-meris \citep[$e.g.$, ][]{2009A&A...507.1675F,2014IPNPR.196C...1F}, from now on we set ET$=$TDB. 

Also, one should correct the reception time for the distance between the antenna receiver and the station clock. 
For a spacecraft light-time solution, the reception time $t_{3R}$ is usually given in Station Time (ST) at the receiving electronics. 
The transformation between ST (usually Coordinated Universal Time, UTC) and ET/TDB is provided in~\citep{2010ITN....36....1P}.
One should then correct for the down-leg delay at the receiver $\delta t_3$ to get the reception time $t_3 (ST)$ at the tracking point of the receiver as
\begin{equation}
    t_{3} = t_{3R} - \delta t_3 \, .
\end{equation}
The same also applies to the emission time $t_1$.
Since one does not know the time of reception and re-transmission $t_2$, the latter is usually determined by iteratively applying Eq.~\eqref{eq:moyerLTit}, $i.e.$, by first considering $t_2 \equiv t_3$ to compute $\rho$, then setting $t_2 \equiv t_3 - \rho$.


\section{Improved light time modeling} \label{sec:improvedmodel}

The standard formulation presented in Section~\ref{sec:standard_dsn} implies an instantaneous retransmission of the signal towards Earth after reception at the spacecraft. In reality, a small delay due to the transponder electronics should be accounted for, which is not provided in the standard auxiliary data. We report in Table~\ref{tab:transpDel} this transponder delay for several probes, as calibrated by industrials on ground before the launch. 

\begin{table}
\noindent \begin{centering}
\def\arraystretch{1.2}
\setlength{\tabcolsep}{9pt}
\begin{tabular}{c|c|c|c}
\multicolumn{1}{c}{Spacecraft} & \multicolumn{1}{c}{Launch} & \multicolumn{1}{c}{TD} & \multicolumn{1}{c}{Source}\tabularnewline
\hline 
MPO 	& 2018	& 4.8-6 & $\dagger$  \tabularnewline
Herschel 	& 2009	& 5.2 & $\dagger$  \tabularnewline
Planck 	& 2009	& 5.2 & $\dagger$  \tabularnewline
MRO & 2005 & 1.4149 & JPL \tabularnewline
Venus Express 	& 2005	& 2.085 & ESOC/FD  \tabularnewline
Messenger 	& 2004				& 1.371 & $\star$ \tabularnewline
Rosetta 	& 2004	& 4.8-6 & $\dagger$  \tabularnewline
Mars Express 	& 2003				& 2.076 & ESOC/FD  \tabularnewline
Mars Odissey &	2001			& 1.4266 & JPL  \tabularnewline
Cassini 	& 1997	& 4.8-6 & $\dagger$  \tabularnewline
MGS &	1996			& 0.7797 & MGS Project  \tabularnewline
\hline 
\end{tabular}
\par\end{centering}
\center \caption{Transponder Delay (TD, $\mu s$) for several probes (MPO $=$ BepiColombo Mercury Planetary Orbiter, MRO $=$ Mars Reconnaissance Orbiter, MGS $=$ Mars Global Surveyor, $\dagger =$~\citet{BussoPriv}, $\star =$~\citet{2007SSRv..131..557S}). \label{tab:transpDel}}
\end{table}
Let us note $\Delta \tau$ this delay in terms of local proper time at the moment and location of the calibration. In our modeling, we have introduced the transponder delay $\delta t_{23}$ in the BCRS between reception and remission events at the probe. In principle, we should relate the calibrated transponder delay $\Delta \tau$ with $\delta t_{23}$. However, the impact of this additional correction shall prove negligible for our purpose, so that in the following $\delta t_{23} \equiv \Delta \tau$.

\subsection{Studied setup}


\par To take into account the transponder delay in the formulation of the light time solution, we need one supplementary event concerning the probe. Let us now consider four events quoted as $\tilde{t}_{l}$. The transmission epoch from Earth is quoted $\tilde{t}_{1}$, $\tilde{t}_2$ is the epoch when the probe received the up-link signal, $\tilde{t}_3$ is the epoch of transmission of the transponded signal towards the Earth and finally $\tilde{t}_4$ is the epoch of reception of the down-link signal at receiving Earth ground station.
We consistently note as ${\bm x}_{\tilde{l}}$ the corresponding position vectors of tracking stations and probe.
The light-time solution is composed of three steps: first we have to determine from the knowledge of the reception event by the Earth receiver the coordinate quantity $\tilde{t}_4-\tilde{t}_3$, then to calculate $\tilde{t}_2-\tilde{t}_1$. The third component deals with the internal electronics delay on-board the probe $\tilde{t}_3-\tilde{t}_2$, {\it i.e.} all kinds of delay between the up-link reception and the down-link emission. 

Our final goal is to express the coordinate quantity $\tilde{\rho}=\tilde{t}_4-\tilde{t}_1$ which is simply $\tilde{t}_4-\tilde{t}_1=(\tilde{t}_4-\tilde{t}_3)+(\tilde{t}_3-\tilde{t}_2)+(\tilde{t}_2-\tilde{t}_1)$. Let us quote the coordinate-dependent quantity $\tilde{t}_3-\tilde{t}_2$ by $\delta t_{23}$. The quantities $\tilde{t}_2-\tilde{t}_1$ and $\tilde{t}_4-\tilde{t}_3$ can be expressed as
\begin{subequations}
	\begin{equation}
		\label{timedelay12}\tilde{t}_2-\tilde{t}_1=\mathcal{T}_r\left(\tilde{t}_2, {\bm x}_{\tilde{1}}, {\bm x}_{\tilde{2}}\right)\, ,
	\end{equation}
and
	\begin{equation}
		\label{timedelay43}\tilde{t}_4-\tilde{t}_3=\mathcal{T}_r\left(\tilde{t}_4, {\bm x}_{\tilde{4}}, {\bm x}_{\tilde{3}}\right)\, ,
	\end{equation}
\end{subequations}
where we used the time transfer functions $\mathcal{T}_r$ introduced in previous publications~\citep{2008CQGra..25n5020T,2014PhRvD..89f4045H}. 
These functions essentially represent the light travel time between two events in a relativistic framework and have an analytical solution at several levels of approximation ($e.g.$,up to the third post-Minkowskian approximation for a static space-time~\citep{2013CQGra..30q5008L} and at the first post-Minkowskian/post-Newtonian approximation for a set of moving axisymmetric bodies~\citep{2014CQGra..31a5021B,2014PhRvD..90h4020H}). 
Hence, at the post-Newtonian level of approximation usually adopted in space navigation for a stationary gravity field, one gets
\begin{equation}
\mathcal{T}_r\left(t_i, {\bm x}_i, {\bm x}_j\right) = \frac{R_{ij}}{c} + RLT_{ij} + \mathcal{O}\left\lbrack c^{-4} \right\rbrack \; . \label{eq:ttf1pn}
\end{equation}


It is then straightforward to 
define the modified round-trip light time $\tilde{\rho}\equiv \tilde{t}_4-\tilde{t}_1+\delta\rho$ as
\begin{equation}
	\tilde{\rho} = \delta t_{23}+\mathcal{T}_r\left(\tilde{t}_2, {\bm x}_{\tilde{2}}, {\bm x}_{\tilde{1}} \right)+\mathcal{T}_r\left(\tilde{t}_4, {\bm x}_{\tilde{4}}, {\bm x}_{\tilde{3}} \right)+\delta\rho  \; , \label{eq:TTFrr}
\end{equation}
where we noted $\delta\rho$ the additional delays ($e.g.$, atmospheric and instrumental delays), supposed equivalent to those given in Eq.~\eqref{eq:moyerLT}.


\subsection{Comparison to the standard formulation}
\par As we have seen in Eq.~\eqref{eq:moyerLT}, the traditional approach used by navigators does not consider the transponder delay in the light time formulation. This results in rewriting Eq.~\eqref{eq:moyerLT} as
\begin{equation}
	\rho = \mathcal{T}_r\left({t}_2, {\bm x}_{{2}}, {\bm x}_{1} \right)+\mathcal{T}_r\left({t}_3, {\bm x}_{{3}}, {\bm x}_{{2}} \right)+\delta\rho  \; , \label{eq:TTFst}
\end{equation}
consisting only in three events $t_1$, $t_2$ and $t_3$. A relation between the $\tilde t_l$ events of our proposed setup and the $t_l$ of the standard setup is easily established by setting \citep[similarly to what proposed in a different context by][]{2002JGeo...34..551D}
\begin{subequations}
\label{eq:tildet} 
\begin{eqnarray}
	\tilde{t_4} &=& t_3  \, , \label{eq:tt4t3}\\
	\tilde{t_3} &=& t_2  \, , \\
   	\tilde{t_2} &=& \tilde{t_3} - \delta t_{23} = t_2 - \delta t_{23} \label{eq:tt2t2}
\end{eqnarray}
and
\begin{eqnarray}
	\tilde{t_1} &=& t_1 - \Delta \rho \; ,
\end{eqnarray}
\end{subequations}
where we used $\Delta \rho = \tilde{\rho} - \rho \equiv (\tilde t_4 - \tilde t_1)-(t_3 - t_1)$ as well as Eq.~\eqref{eq:tt4t3}. As a consequence, we also get
\begin{subequations}
\label{eq:tildex} 
\begin{eqnarray}
	\bm x_{\tilde{2}} (\tilde t_2) &=& \bm x_2 (t_2 - \delta t_{23})  \, , \\
	\bm x_{\tilde{1}} (\tilde t_1) &=& \bm x_1 (t_1 - \Delta \rho) \, .
\end{eqnarray}
\end{subequations}


Thus, it is straightforward to analyze the difference between Eq.~\eqref{eq:TTFrr} and Eq.~\eqref{eq:TTFst}. 
Since the transponder delay $\delta t_{23}$ is roughly equal to several $\mu$s (see Table~\ref{tab:transpDel}), we perform a Taylor expansion of Eq.~\eqref{eq:TTFrr} and we introduce Eqs.\eqref{eq:tildet}-\eqref{eq:tildex}, such that
\begin{eqnarray} \label{eq:extratermTTF}
\tilde{\rho}&=&\mathcal{T}_r\left(t_2-\delta t_{23}, {\bm x}_{{2}} - \bm v_2 \delta t_{23}, {\bm x}_{{1}}  - \bm v_1 \Delta \rho \right)+\delta t_{23} \nonumber\\
			&& + \mathcal{T}_r\left(t_3,{\bm x}_{3},{\bm x}_{2}\right) +\delta\rho \nonumber \\
			&=&\mathcal{T}_r\left(t_2, {\bm x}_2, {\bm x}_1\right)+\delta t_{23}+\mathcal{T}_r\left(t_3,{\bm x}_3,{\bm x}_2\right) +\delta\rho \nonumber\\
            && - \delta t_{23} \, \frac{\partial\mathcal{T}_r\left(t, {\bm x_2},{\bm x_1}\right)}{\partial t}\Big\vert_{{\bm t}={\bm t}_2}  \nonumber \\
		&  & - \delta t_{23} \, v_2^i \frac{\partial\mathcal{T}_r\left(t_2, {\bm x},{\bm x_1}\right)}{\partial x^i}\Big\vert_{{\bm x}={\bm x}_2}  \nonumber \\
		&& - \Delta \rho \, v_1^i \frac{\partial\mathcal{T}_r\left(t_2, {\bm x}_2,{\bm x}\right)}{\partial x^i}\Big\vert_{{\bm x}={\bm x}_1} +\mathcal{O}\left\lbrack(\delta t_{23}, \Delta \rho)^2\right\rbrack\nonumber \\
		&\equiv & \rho - \delta t_{23} \, \frac{\partial\mathcal{T}_r\left(t, {\bm x_2},{\bm x_1}\right)}{\partial t}\Big\vert_{{\bm t}={\bm t}_2}  \nonumber \\
		&  & - \delta t_{23} \, v_2^i \frac{\partial\mathcal{T}_r\left(t_2, {\bm x},{\bm x_1}\right)}{\partial x^i}\Big\vert_{{\bm x}={\bm x}_2} \\ 
		&& - \Delta \rho \, v_1^i \frac{\partial\mathcal{T}_r\left(t_2, {\bm x}_2,{\bm x}\right)}{\partial x^i}\Big\vert_{{\bm x}={\bm x}_1} +\mathcal{O}\left\lbrack(\delta t_{23}, \Delta \rho)^2\right\rbrack \, ,\nonumber    
\end{eqnarray}  
where ${\bm v}_l=\lbrace v_l^i\rbrace$ is the coordinate velocity of the probe at instant $t_l$. It is worth noting that since an analytical formulations of the time transfer function $\mathcal{T}_r$ is available at various level of approximation, Eq.\eqref{eq:extratermTTF} can be easily adapted for increasing accuracies.  
For this application it is sufficient to expand $\mathcal{T}_r$ using Eq.~\eqref{eq:ttf1pn}, which finally gives
\begin{equation}
	\Delta \rho = \tilde{\rho} - {\rho} = \delta t_{23} \left( 1 + \frac{(\bm v_1 - \bm v_2) \cdot \bN_{12}}{c} \right) +\mathcal{O}\left\lbrack(\delta t_{23})^2, c^{-2}\right\rbrack \;  \label{eq:rhotilde}
\end{equation}
with
\begin{equation}
\bN_{12} \equiv \frac{\bx_2-\bx_1}{\norm{\bx_2-\bx_1}} \; \nn .
\end{equation} 
While the constant term $\delta t_{23}$ is usually calibrated in the computed Range, Eq.~\eqref{eq:rhotilde} highlights the presence of an extra non-constant term, directly proportional to the transponder delay and neglected in Moyer's model. This term also depends on the position and velocity of both the probe and the ground station. 

It can be  physically interpreted as a modification of the determination of the state vector at transponding event of coordinate time $t_2$ or as an imprecise determination of the time $t_2$. Both range and Doppler are then affected by this mismodeling, as we show in Section~\ref{sec:observables}.

 





\section{Impact on Range and Doppler computed observables} \label{sec:observables}

Based on the standard and modified formulation of the light time $\rho$ and $\tilde \rho$, respectively, we derive additional terms appearing in Range and Doppler observables.

\par The computed Range Observable $R$ is simply given by
\begin{equation}
    R = K \rho \, ,
\end{equation}
where $K$ is a conversion factor. Depending on the processing strategies, the transponder delay $\delta t_{23}$ is either added to $\rho$ or estimated together with other error sources in a so called "range bias". However, both these solutions do not fully account for the impact of the transponder delay as given by Eq.~\eqref{eq:rhotilde}, in particular regarding the time dependent terms.




Regarding Doppler, the basic idea is to measure the frequency shift based on the emission and reception times of a series of signals over a given time interval. Several configurations are possible.
Two and three-way Doppler (in the latter the signal is emitted and received by different stations) are usually ramped, meaning that the emitted frequency $f_T$ changes with time following a piecewise linear function of time. For our purpose, we consider a simple modeling of unramped two-way Doppler $F_2$, such that $\dfrac{\partial f_T}{\partial t}=0$.
Hence,
\begin{equation}
	F_2 = \frac{M_2 f_T}{T_c} (\rho_e - \rho_s) \; , \label{eq:F2}
\end{equation}
where $M_2$ is a multiplying factor related to the transponded frequency and $\rho_e$ and $\rho_s$ are the light-times of two signals whose receptions are separated by a "counting time" $T_c$, typically of the order of $10-60$ s.

The difference between computing a Doppler observable with the two formulations presented in this paper is then given by introducing Eq.~\eqref{eq:rhotilde} into Eq.~\eqref{eq:F2} as
\begin{eqnarray} \label{eq:DF2}
	\Delta F_2 &=& \tilde F_2 - F_2 \nonumber \\
   	           &=& \frac{M_2 f_T}{T_c} \left\lbrack(\tilde\rho_e - \tilde\rho_s)-(\rho_e - \rho_s) \right\rbrack  \\
	           &=& \frac{M_2 f_T}{T_c} \frac{\delta t_{23}}{c} 
               \left\lbrack (\bm v_1^e - \bm v_2^e) \cdot \bN_{12}^e - (\bm v_1^s - \bm v_2^s) \cdot \bN_{12}^s \right\rbrack \; . \nonumber 
\end{eqnarray}
The transponder delay $\delta t_{23}$ itself is simplified when differencing, but not its impact on the Doppler frequency. Indeed, the epochs at which both the spacecraft and ground station positions are evaluated in the uplink change.












\section{Numerical applications} \label{sec:applications}
In this section we present some examples to analyze the impact of the transponder delay $\delta t_{23}$ in some realistic configurations.
First, we compute the time dependent terms given in Eqs.~\eqref{eq:rhotilde} and~\eqref{eq:DF2} during the Earth flyby of several probes. Then, we show how the transponder delay can be easily introduced in the processing pipeline of Doppler data by explicitly adding a constant $\delta t_{23}$ to the light-time algorithm as in Eq.~\eqref{eq:tt2t2}, thus retrieving the probe trajectory at a (slightly) different epoch.
We perform the latter test on the GRAIL and BepiColombo missions within the planetary extension of the Bernese GNSS Software \citep[BSW,][]{bernese.book}, mainly based on~\citet{Moyer:2000} for the computation of deep space observables~\citep{issfd2015_bertone}.

\subsection{Application to Earth flybys}


In order to evaluate the magnitude of the additional term in Eq.~\eqref{eq:rhotilde}, we compute $\Delta \rho = \tilde \rho - \rho$ and its time derivative $\Delta \dot \rho =  \dot{\tilde\rho} - \dot\rho = \dfrac{\Delta F_2}{M_2 f_T}$, $i.e.$, the difference between the range-rate calculated by the two models. 
We consider several probes (Rosetta, NEAR, Cassini, Galileo) during their Earth flyby, which is a particularly favorable configuration thanks to the quick changes in the relative velocity vector between probe and antenna. Also, close approaches are an important source of information when measuring the geophysical parameters of a celestial body. We use the NAIF/SPICE toolkit~\citep{2011epsc.conf...32A} to retrieve the ephemeris for probes and planets to be used in the computation.

\begin{figure}[ht] 
	\centering
	\includegraphics[width=0.78\linewidth]{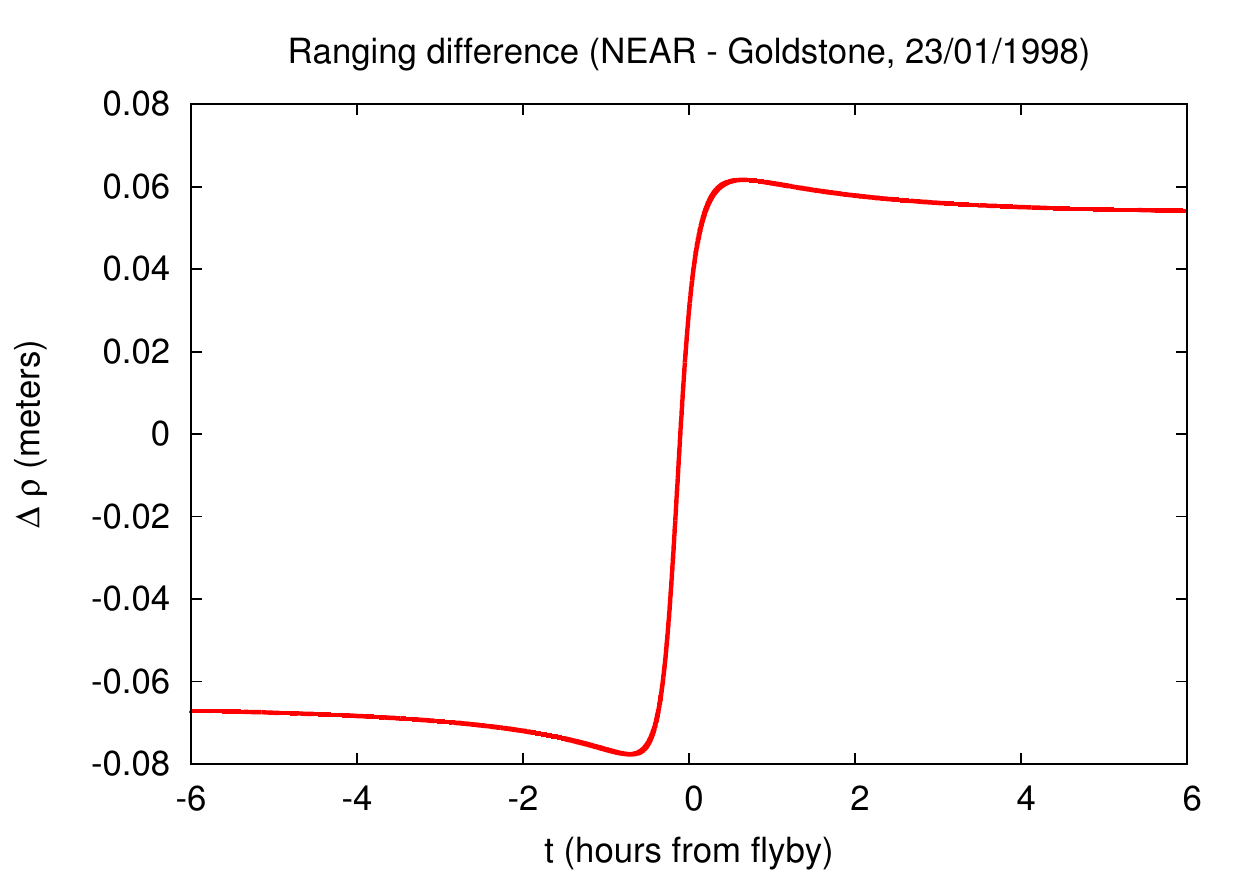}
	\caption{Light time difference $\Delta \rho$ (meters - hours from flyby) during NEAR Earth flyby.}
	\label{fig:range}
\end{figure}

\begin{figure}[ht] 
	\centering
	\includegraphics[width=0.78\linewidth]{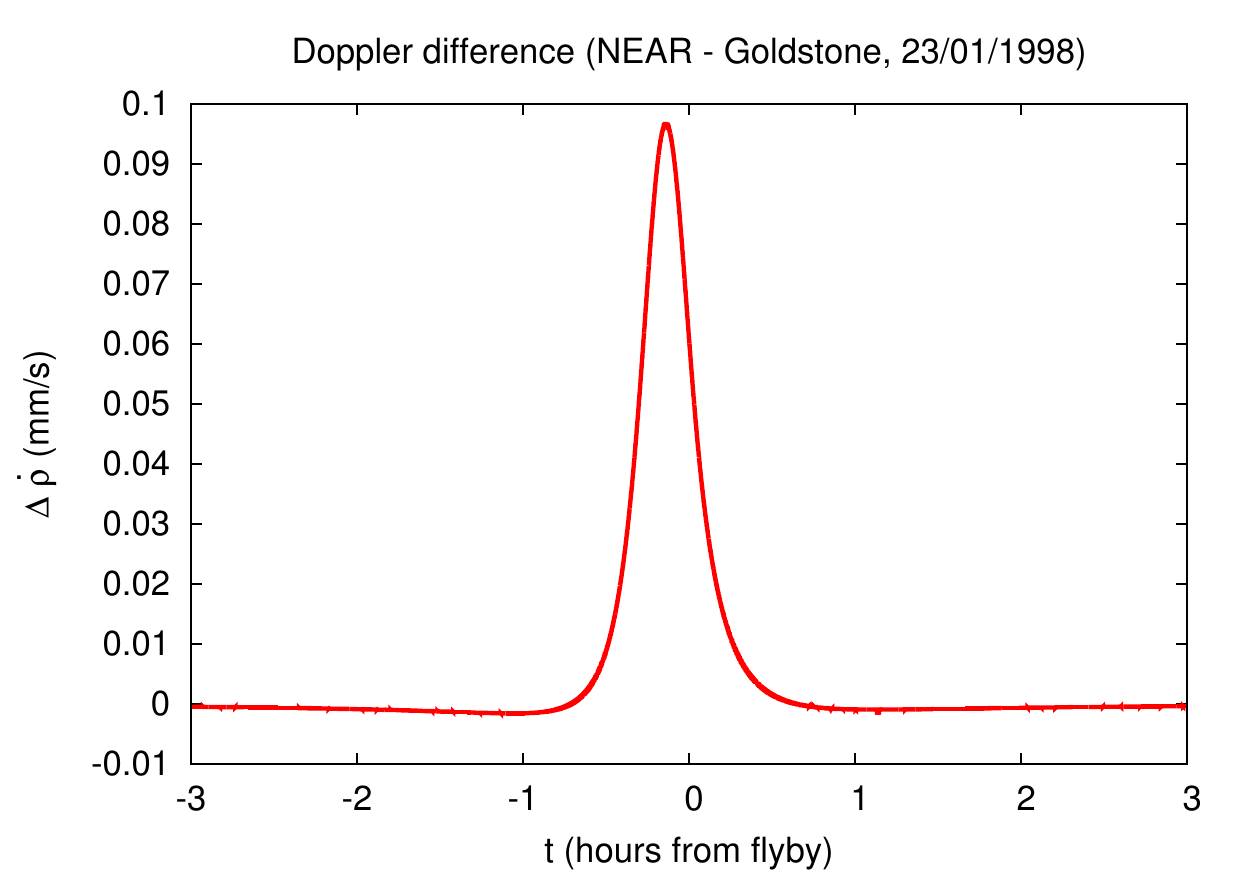}
	\caption{Range-rate difference $\Delta \dot \rho$ (mm/s - hours from flyby) during NEAR Earth flyby.}
	\label{fig:dop}
\end{figure}

We fix $\delta t_{23} = 10~\mu$s and compute Eq.~\eqref{eq:rhotilde} and its time derivative from Eq.~\eqref{eq:DF2} for the NEAR probe during its Earth flyby on 23 January 1998. We find a difference of the order of some $cm$ for the probe distance $c \Delta \rho$ calculated by the two models (when subtracting the constant $\delta t_{23}$ bias) and a difference up { to several $10^{-2} \; mm/s$} at the instant of maximum approach {for its velocity}. These results are shown in Figures~\ref{fig:range} and~\ref{fig:dop} for $T_c = 1$~s. We note that changing the integration time $T_c$ only has a significant impact when it becomes larger than several minutes. Also, results for other $\delta t_{23}$ values can be easily deduced as $\Delta \rho$ and $\Delta \dot \rho$ are directly proportional to the transponder delay. The amplitude of such effects are in principle within the nominal accuracy of future missions expected to perform Earth gravity-assist maneuvers, such as BepiColombo.\\

\begin{figure}[ht]
	\centering
	\includegraphics[width=0.78\linewidth]{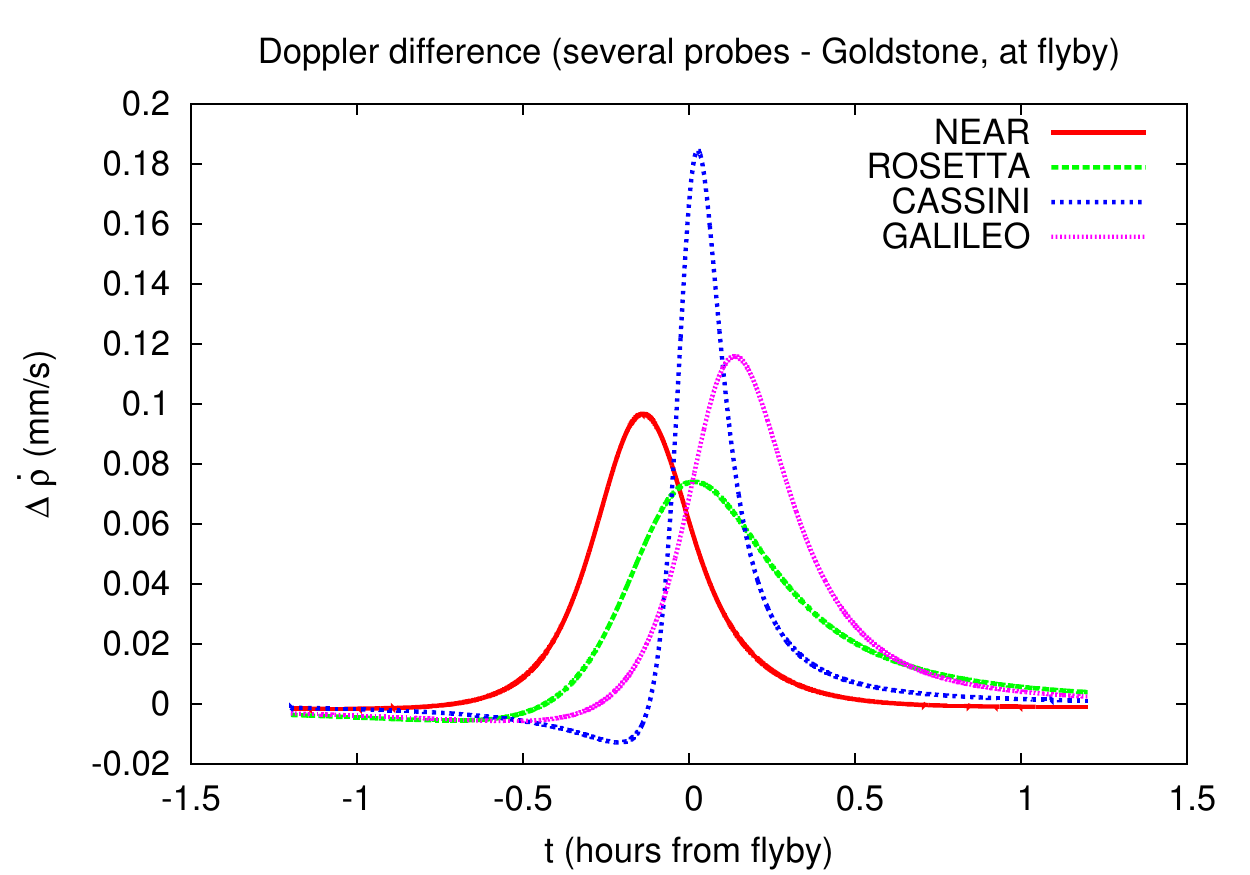}
	\caption{Doppler difference $\Delta \dot \rho$ (mm/s - hours from Earth flyby) for several probes with respect to Goldstone DSN station. The results highlight the high variability of the effect on Doppler measurements.}
	\label{fig:dop1}
\end{figure}

In order to highlight the high variability of the transponder delay effect on Doppler {measurements}, we also compute $\Delta \dot \rho$ for different probes in different configurations with respect to the observing station. The results displayed in Figure~\ref{fig:dop1} show that this delay cannot be simply calibrated by adding a constant Range bias and hint that it should be carefully dealt with for the Doppler computation.

A preliminary study presented in~\citet{2013arXiv1305.1950B} compared the results of this section with the so called flyby anomaly \citep{2008PhRvL.100i1102A} but found the mismodeling of the trans\-ponder delay to possibly account only for a few percent of it. 

\subsection{Application to GRAIL Doppler and KBRR data}

Here, we use the BSW to process two-way S-band Doppler data and to retrieve GRAIL-A and GRAIL-B orbits around the Moon for several days of the primary mission phase. In particular, we selected both days when the orbital plane was parallel (days 63-64 of year 2012) and when it was perpendicular (days 72-73 of year 2012) w.r.t. the line of sight between the satellites and the Earth. We fit a set of 6 orbital elements
in daily arcs using GRGM900C~\citep{2014GeoRL..41.3382L} up to d/o 600 as background gravity field.
We first use the standard modeling for light time and Doppler and then compute alternative orbits by adding an arbitrary transponder delay of $2.5~\mu$s to the light-time computations. The 
resulting orbit differences for the two satellites are shown in Table~\ref{tb:GRAdiff} and are well within the uncertainty value of the orbit recovery (estimated in several cm in radial direction and $\approx 1$ m in the other directions).


\begin{table}[ht]
\centering
\begin{tabular}{lcccc}
 DOY & GRAIL & Radial & Along-Track & Cross-Track  \\
\hline
\hline
12-063 & A & 0.08 & 0.62 & 2.96 \\
	   & B & 0.11 & 1.43 & 1.12 \\
\hline
12-064 & A & 0.10 & 1.79 & 7.29 \\
	   & B & 0.10 & 0.67 & 2.72 \\
\hline
\hline
12-072 & A & 0.08 & 1.11 & 0.10 \\
	   & B & 0.18 & 0.60 & 0.14 \\
\hline
12-073 & A & 0.04 & 0.16 & 0.45 \\
	   & B & 0.04 & 1.01 & 0.87 \\
\end{tabular}
\caption{Orbit differences (mm, orbit frame) caused by introducing the transponder delay in the Doppler modeling for the orbit improvement process. During Day of Year (DOY) 12-063/064 the orbital plane of the GRAIL satellites is parallel to the line of sight w.r.t. Earth, while it is perpendicular for DOY 12-072/073.}
\label{tb:GRAdiff}
\end{table}

Based on both orbit pairs, we then compute the Ka-band inter-satellite Range-Rate (KBRR), $i.e.$ the radial velocity along the line-of-sight between the two satellites. 
Their difference shows a once-per-revolution signal with an amplitude of $\approx 0.1-1\;\mu$m/s in the along-track direction due to the mismodeling of the transponder delay.
For completeness and to evaluate the impact of the transponder delay on the operative orbit recovery of the GRAIL probes, we perform a further orbit improvement by fitting both pair of orbits to Doppler and KBRR data. The relative weighting of these observables is usually chosen to strongly favor KBRR data (here we apply a $1:10^8$ ratio) because of their higher accuracy. A comparison of the resulting orbits then shows that post-fit KBRR differences due to the transponder delay are reduced to $\approx 0.001\; \mu$m/s (to be compared with the nominal KBRR accuracy of $0.03\; \mu$m/s). KBRR residuals result globally improved by our updated light-time algorithm, but well below the formal uncertainties.


\subsection{Application to ESA BepiColombo mission}

\begin{figure}[ht]
	\centering
	\includegraphics[width=0.8\linewidth]{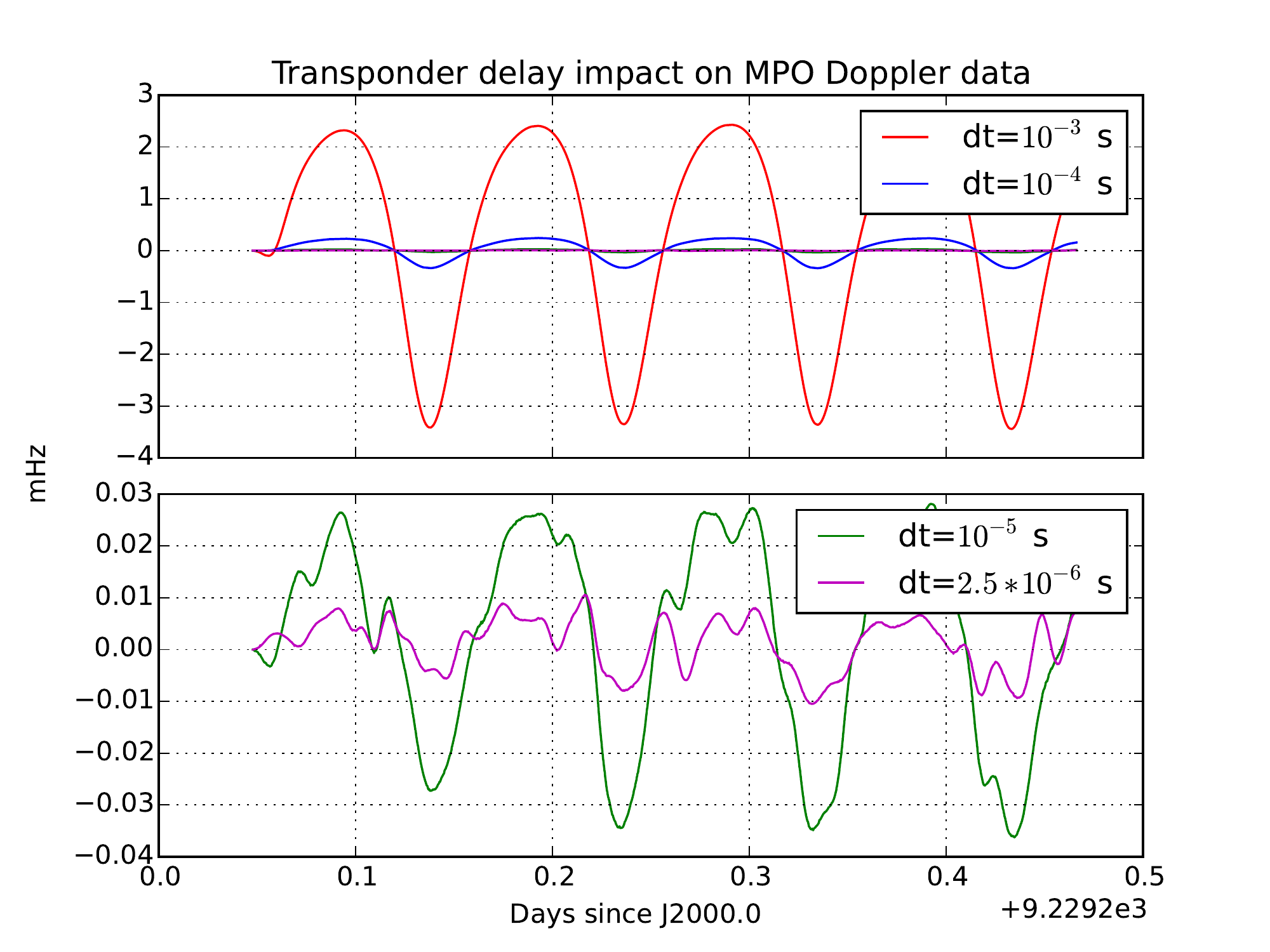}
	\caption{Doppler difference $\Delta \dot \rho$ for the nominal MPO orbit around Mercury on 08/04/2025 for different values of the transponder delay $\delta t$ ($1 $ mHz $\approx 0.035 $ mm/s @ $8.4$ GHz).}
	\label{fig:MPOdtRes}
\end{figure}

We use the BSW to simulate two-way X-band Doppler for BepiColombo Mercury Planetary Orbiter (MPO) nominal orbit retrieved from ESA Spice SPK for 08/04/2025.
We first compute Doppler data as observed by the Deep Space Network antennas following the standard formulation by~\citet{Moyer:2000}. Then, we include the transponder delay in the light-time modeling used for the simulation. We compute the resulting Doppler signal for several values of $\delta t_{23}$ in the range $10^{-6}-10^{-3}$ s and show the differences w.r.t. the standard formulation in Fig.~\ref{fig:MPOdtRes}. As shown in Table~\ref{tab:transpDel}, MPO transponder delay has been measured at $4.8-6~\mu$s.

Our results highlight an additional frequency signal superposed to the orbital period and showing an amplitude linearly dependent from $\delta t_{23}$, as expected from Eq.~\eqref{eq:rhotilde}. The amplitude of the additional signal, neglected in the standard formulation, is up to  several mHz for slow transponders ($\delta t \approx 1$ ms) while it accounts for $\approx 0.02$ mHz for modern transponders ($\delta t \approx 2.5~\mu$s). These values should be compared to the nominal accuracy of the MORE instrument~\citep{2001P&SS...49.1597I}, which is $\approx 5$ mHz and $\approx 1.5$ mHz at $10$ seconds integration time for X- and Ka-bands, respectively~\citep{2016MNRAS.457.1507C}.
The impact of the transponder delay looks then safely below the noise level for the BepiColombo mission in its science phase.

\section{Conclusions} \label{sec:conclusions}

In this communication, we present a refinement of the formulation of two-way light-time for the tracking of space probes. In particular, we focus on the transponder delay, a tiny delay (amounting to several $\mu$s in modern devices) between the reception of the signal on the spacecraft and its re-emission towards Earth. 
It seems obvious from our results that the influence of the transponder delay cannot be reduced to a simple correction with a constant bias without taking some precautions. It is {indeed} responsible for a tiny effect on the computation of light time and has an impact on both range and {Doppler} determination. We take it into account by a more complete modeling, considering four events in the observables modeling instead of three as in Moyer. 

In order to test the amplitude and variability of this effect on real data, we compute its influence on some real probe-ground station configurations during recent Earth flybys (NEAR, Roset\-ta, Cassini and Galileo).
The observables calculated using the standard model and our updated one show differences of the order of several cm and of $0.1$ mm/s for the range and the {range-rate}, respectively.
As expected from our analytical results, the impact of the transponder delay is maximized during a flyby maneuver, when the relative velocity between spacecraft and observer changes rapidly. Nevertheless, as already shown in~\citet{2013arXiv1305.1950B}, this effect can only account for a tiny portion of the so-called flyby anomaly. 
Moreover, we use the planetary extension of the Bernese GNSS Software to simulate the impact of several amplitudes of the transponder delay on both Doppler data and orbit recovery for the NASA GRAIL and ESA BepiColombo missions. To do so, we modify the light-time computation algorithm for the up-leg by requesting the probe ephemeris at an epoch anticipated of $\delta t_{23}$. 
The highlighted differences are acceptable for most operational goals at present, although applying a more accurate modeling could avoid the possible propagation of orbital errors in, $e.g.$ the recovery of geophysical signatures or the analysis of tiny relativistic signals~\citep{2007AcAau..61..932M} which could correlate with the effects of the transponder delay.
Also, since the MORE instrument on-board BepiColombo will be equipped with an internal calibration circuit, it will be possible to measure the transponder delay and to systematically apply the updated formulation provided in this paper to test the impact on the data processing.

Finally, we stress that this error is directly proportional to the transponder delay. This means that this effect might be relevant for past missions equipped with slower transponders (whose data are still largely used for scientific purposes) or for long lasting missions when considering the degraded performances of aging transponders.
In the future too, increasing spacecraft tracking accuracy~\citep{2014AcAau..94..699I} should be accompanied by the development of faster transponders or by correctly measuring, distributing and accounting for this delay in the orbit determination process.

\section{Acknowledgments}
\noindent SB acknowledges the financial support of the Swiss National Science Foundation (SNF) via the NCCR PlanetS. SB also thanks Dr.~A. Hees for the helpful discussions. The authors thank G.~Canepa and A.~Busso of Thales Alenia Space for providing information about the transponder delay of several probes.\\


\bibliographystyle{model2-names.bst}\biboptions{authoryear}
\bibliography{BiblioCNAP}

\end{document}